\documentclass[preprint,prc,showpacs,showkeys,longbibliography]{revtex4-1}
\usepackage{epsfig,graphicx,float,color}
\usepackage{amssymb}
\usepackage[normalem]{ulem}
\begin{document}

\title{Two-neutron transfer reactions as a tool to study the interplay between shape coexistence and quantum phase transitions}

\author{J.E.~Garc\'{\i}a-Ramos$^{1,2}$, J.M. Arias$^{2,3}$, and A.~Vitturi$^4$}
\affiliation{
$^1$Departamento de  Ciencias Integradas y Centro de Estudios Avanzados en F\'isica, Matem\'atica y Computaci\'on, Universidad de Huelva,
21071 Huelva, Spain\\
$^2$Instituto Carlos I de F\'{\i}sica Te\'orica y Computacional,  Universidad de Granada, Fuentenueva s/n, 18071 Granada, Spain\\
$^3$Departamento de F\'{\i}sica At\'omica, Molecular y Nuclear, Universidad de Sevilla, Apdo.\ 1065, 41080 Sevilla, Spain \\
$^4$Dipartimento di Fisica e Astronomia ``Galileo Galilei'', Universit\'a degli Studi di Padova, Padova, Italy
} 
\begin{abstract} 
  The atomic mass table presents zones where the structure of the states changes rapidly as a function of the neutron or proton number. Among them, notable examples are the $A\approx 100$ Zr region, the Pb region around $N=104$ neutron midshell or the $N\approx 90$ rare-earth region. The observed phenomena can be understood in terms either of shape coexistence or quantum phase transitions.
The goal of this study is to find an observable that could distinguish between both phenomena, shape coexistence and quantum phase transitions. The selected observable to be analyzed is the two-neutron transfer intensity between the $0^+$ states in the parent and daughter nuclei. The framework in which the study is done is the Interacting Boson Model (IBM), including its version with configuration mixing (IBM-CM). In order to generate the wave functions of the isotope chains of interest, needed for calculating transfer intensities, previous systematic studies with IBM and IBM-CM are taken without changing the parameters. 
Results for two-neutron transfer intensities are presented for Zr, Hg and Pt isotopic chains using IBM-CM and, moreover, the same is done for Zr, Pt and Sm isotopic chains using IBM with just a single configuration, i.e., without using configuration mixing.
In the case of Zr, the two-neutron transfer intensities between the ground states provide a clear observable indicating that normal and intruder configurations coexist in the low-lying spectrum and that they cross at $A=98 \rightarrow 100$,
and this could allow to disentangle whether or not shape coexistence is inducing a given QPT.
In the case of Pt, where shape coexistence is present and the regular and the intruder configurations cross for the ground state, there is almost no influence in the value of the two-neutron transfer, neither in the case of Hg where the ground state always has regular nature. For the Sm isotope chain that is one of the quantum phase transition paradigms, the value of the two-neutron transfer is strongly affected.
\end{abstract}
 

\keywords{Two-nucleon transfer reactions, shape coexistence, quantum phase transition, interacting boson model.}

\date{\today}
\maketitle

\section{Introduction}
\label{sec-intro}

A quantum phase transition (QPT) implies an abrupt change of the ground state properties of the system under study when a control parameter reaches a critical value \cite{Cejn09,Cejn10,Sach11}. In the case of Nuclear Physics, usually, a complete chain of isotopes is studied in order to see systematic variations and, eventually, abrupt changes in the ground state properties \cite{diep80a}. In this case the control parameter is the neutron number. Typically, QPTs are found in transitional regions where the structure of the nuclei evolves in between two different limiting structures (symmetries), either spherical and rigidly deformed, or spherical and gamma-unstable or gamma-unstable and rigidly deformed. The appearance of a QPT implies and, consequently, has been characterised by peculiar changes in certain observables. Just to cite few of them:
\begin{itemize}
\item The slope of the two-neutron separation energy presents a discontinuity.
\item There is a minimum in the excitation energy of certain states such as the $0_2^+$ state and an increase in the density of states.
\item There is a sudden drop down in the excitation energy of certain states such as the $2_1^+$ state.
\item There is a rapid increase in B(E2: $2_1^+\rightarrow 0_1^+$), among other transition probabilities. 
\item There is a rapid increase of the  $2_1^+$ quadrupole moment.
\end{itemize}
Moreover, for nuclei that are placed at the critical point, Iachello introduced the concept of critical point symmetry \cite{Iach00,Iach01,Iach03}, which provides values of energy ratios and transition probabilities that are parameter free.  

Shape coexistence is a very broad phenomenon that appears almost everywhere in the nuclear mass table but more specifically near the proton or neutron shell closures \cite{wood92,heyde11}. It supposes the presence of states with very different shapes or deformation, for instance vibrational-like and deformed, in a narrow excitation energy range. The existence of different configurations is associated with particle-hole (np-nh) excitations across the shell closure. Typically, vibrational-like states correspond to 0p-0h excitations while the deformed ones are associated to 2p-2h excitations. At this point it is worth to remember that the shape of the given state is not an observable, though is an extensively used concept, but can be extracted from the Kumar-Cline sum-rule  over E2 matrix elements \cite{kumar72,Cline86}. The presence of shape coexistence has a strong influence on the spectroscopic properties of a chain of nuclei:
\begin{itemize}
\item There is a family of states with a clear parabolic-like excitation energy systematics centered at the midshell when depicted as a function of the neutron (proton) number.
\item The crossing of states belonging to different families implies sudden changes in the deformation of the states. These crosses lie before and after midshell.
\item There is a drop in the energy of a number of $0^+$ states. In general, several $0^+$ states sit very low in energy.
\item When both families of states cross in the ground state, it experiences an abrupt change of deformation with consequences in the systematics of the two-neutron separation energy, the quadrupole moment or the B(E2: $2_1^+\rightarrow 0_1^+$) values.
\end{itemize}

According to the above list, QPT and shape coexistence show similar systematics and in many cases it is not simple to disentangle which one is the \emph{responsible} of the rapid onset of deformation. This work is focused in the study of a different quantity such as the two-neutron transfer intensity to analyze whether it can provide us with an observable that displays a different behaviour under the presence of a QPT, shape coexistence or a QPT induced by shape coexistence. This tool was first used to explore the existence of QPTs in Ref.\cite{Foss07}, later in \cite{Zhan17} and more recently in \cite{Nomu19}. Moreover, the two-neutron transfer intensity has been used to analyze the possible interplay between QPT and shape coexistence in \cite{Vitt18,Vittu18a} proving that some differences are in order when comparing a schematic QPT with a schematic shape coexistence situation, namely, the two-neutron transfer intensity is much more fragmented in the case of a QPT than in the case of the presence of shape coexistence, though in both cases  rapid changes appear in this observable. In this work we will explore in detail a set of realistic calculations. 

Several regions are of great interest. Zr region is characterized by very rapid changes in the ground state structure. In particular, the onset of deformation when passing from $^{98}$Zr to $^{100}$Zr is one of the fastest ever observed in the nuclear chart. It has been probed both experimental and theoretically that certain excited states of Zr isotopes have different shapes than the ground state \cite{Togashi16}. This change is associated to low-lying intruder configurations. In the case of Pb region, the evolution of the structure of the ground state is quite stable but the spectra own states with different shapes, i.e., shape coexistence is present, as in Pb, Hg and Pt nuclei \cite{heyde11}. On the other hand, in the rare-earth region, changes in the structure of the ground state have been associated to a QPT \cite{Garc03}, for instance in the Sm isotope chain \cite{Iach98}.

\begin{figure}[htb]
\centering
\includegraphics[width=.9\linewidth]{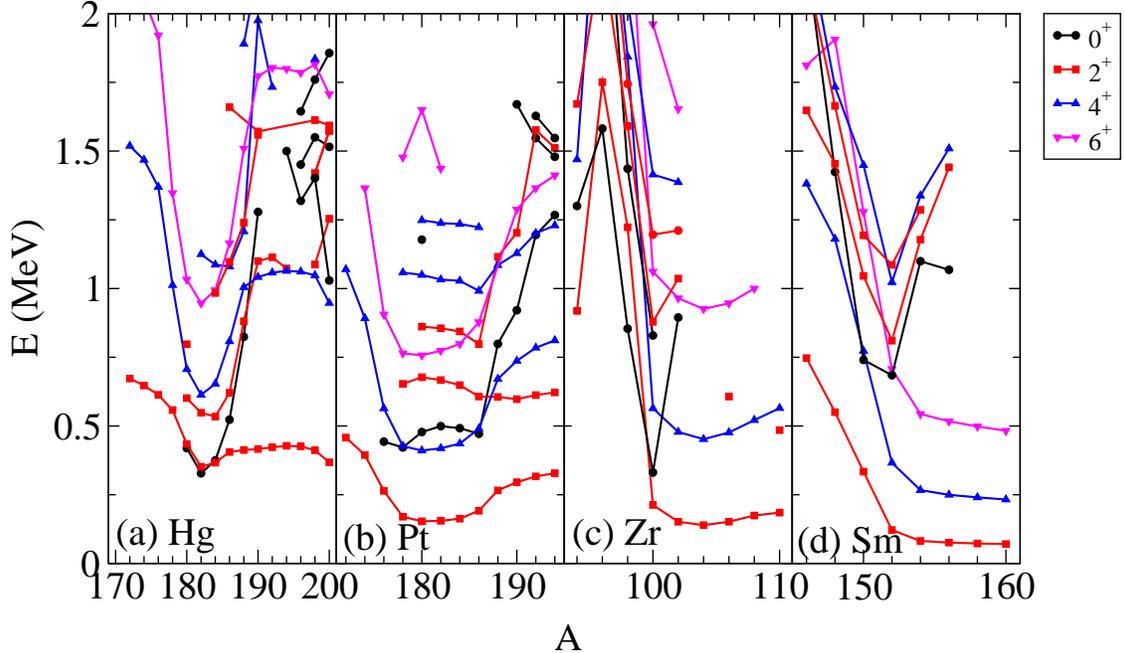}
\caption{Partial spectra of Hg (panel (a)), Pt (panel (b)), Zr (panel (c)), and Sm isotopes (panel (d)).}
\label{fig:spectra-hg-pt-zr-sm}
\end{figure}

Although the onset of a QPT or the importance of shape coexistence have been studied in detail for the aforementioned isotopes in a large set of publications, we present here, for completeness, the partial spectra of Hg, Pt, Zr, and Sm isotopes in Fig.~\ref{fig:spectra-hg-pt-zr-sm}. In this figure, the spectra of Hg isotopes clearly illustrate the existence of a family of intruder states with a parabolic trend, centered at the midshell, $N=104$ \cite{Garc14b}. In the case of Pt the presence of intruder states is not so clear and it is said that shape coexistence is concealed with a not so clear parabolic like energy systematics for the intruder states \cite{Garc09,Garc11}, although the energy trend is still symmetric with respect to the midshell. For Zr isotopes it is observed the sudden onset of deformation, as can be deduced from the low energy of the $2_1^+$ state and of the yrast band, in general, and that can be interpreted as the existence of a QPT \cite{Togashi16}, moreover, it is remarkable the very low energy of the $0_2^+$ state and the influence of the midshell, $N=66$, on the energy systematics of the yrast band, which suggests the presence of configuration mixing \cite{Garc19}. Finally, for Sm isotopes it is also observed the rapid onset of deformation, with the lowering of the energy of the yrast band and the existence of a QPT \cite{Garc03}, with also  a notable decreasing in energy of the $0_2^+$ state, although not so much as for the Zr case, moreover, there are no hints pointing to the existence of shape coexistence.

The concept of QPT was also analyzed in the framework of IBM-CM, first of all defining a method to construct an energy functional considering different configurations, as introduced in \cite{Frank02,Frank04} and then through the construction of a schematic phase diagram of the model \cite{Frank06}. Later on, the phase diagram of the IBM-CM Hamiltonian was studied more in detail in \cite{Hell07,Hell09}. The phase diagram in these cases is much more involved than for a single configuration and it is hard to obtain a single phase diagram considering all possible parameters.    

The paper is organised as follows. In Section \ref{sec:formalism} the formalism for calculating two-nucleon transfer intensities is briefly revisited. In Section \ref{sec:two-nucleon-transfer} a systematic study of the two-nucleon transfer intensities for several isotope chains is presented, first considering configuration mixing calculations (subsection \ref{sec:resultsI}) and then using a single configuration (subsection \ref{sec:resultsII}). Finally, our main conclusions are listed in Section \ref{sec:conclusions}.

\section{The formalism}
\label{sec:formalism}
In order to compute two-nucleon transfer intensities we will use here as framework the IBM \cite{iach87}. The model was proposed as a symmetry dictated approximation of the shell model, assuming that the relevant degrees of freedom are pairs of nucleons coupled either to angular momentum $L=0$ (S pairs) or to angular momentum $L=2$ (D pairs), which are considered as bosons as further approximation. The number of active pairs of nucleons regardless its particle or hole nature or proton or neutron character is denoted as $N$. Very soon the IBM was modified to deal also with particle-hole excitation, e.g., 2p-2h excitations \cite{duval81,duval82}. In this case the original Hilbert space is enlarged to $[N]\oplus[N+2]$. The $[N+2]$ space corresponds to considering two extra bosons that come from the promotion of a pair of protons across the corresponding shell closure, generating an extra boson made of proton holes and another made of proton particles. 

For the case of the IBM without configuration mixing a simplified Hamiltonian called extended consistent-Q Hamiltonian (ECQF) \cite{warner83,lipas85} will be considered. This Hamiltonian has been used to describe successfully even-even medium- and heavy-mass nuclei. The Hamiltonian can be written as
\begin{equation}
  \hat{H}_{\rm ecqf}=\varepsilon \hat{n}_d+\kappa' \hat{L}\cdot\hat{L}+  \kappa \hat{Q}\cdot\hat{Q},
  \label{eq:cqfhamiltonian}
\end{equation}
where the operators appearing in the Hamiltonian are, respectively, the $d$ boson number, the angular momentum and the quadrupole operator. $\varepsilon, \kappa'$ and $ \kappa$ are parameters of the model.

The considered Hamiltonian for the case of IBM-CM  has been
\begin{equation}
  \hat{H}=\hat{P}^{\dag}_{N}\hat{H}^N_{\rm ecqf}\hat{P}_{N}+
  \hat{P}^{\dag}_{N+2}\left(\hat{H}^{N+2}_{\rm ecqf}+
    \Delta^{N+2}\right)\hat{P}_{N+2}\
  +\hat{V}_{\rm mix}^{N,N+2}~,
\label{eq:ibmhamiltonian}
\end{equation}
where $\hat{H}^N_{\rm ecqf}$ and  $\hat{H}^{N+2}_{\rm ecqf}$ correspond to the ECQF Hamiltonians (\ref{eq:cqfhamiltonian}) for the regular and intruder sectors, $\hat{P}_{N}$ and $\hat{P}_{N+2}$ are projection operators onto the $[N]$ and the $[N+2]$ boson spaces, respectively, $\hat{V}_{\rm mix}^{N,N+2}$  is describing the mixing between the $[N]$ and the $[N+2]$ boson subspaces, and  the parameter $\Delta^{N+2}$ accounts for the energy needed to promote a pair of protons across the proton shell closure \cite{Hey85,Hey87}.

Once the parameters of the Hamiltonian have been fixed, the wave functions are available and they can be used to calculate the value of the two-neutron transfer intensities. Since we will be interested just in two-neutron transfer reactions between the low-lying $L^\pi=0^+$ states, for the transfer operator we will use the simplest possible option, namely, $\hat {\cal P}^\dagger = \hat s^\dagger$ and $\hat {\cal P} = \hat s$, being the transfer intensities  proportional to the square of the reduced matrix element. Therefore, we implicitly assume that the results do not strongly depend on the precise structure of the transfer operator. In the case of the IBM-CM the intensities can be written as (note that the emission or absorption of a nucleon pair implies that the number of bosons changes in one unit), 
\begin{equation}
  \label{eq:op_pt}
 I(N,0_i^+\rightarrow N-1, 0_f^+)= \left|\langle N-1, L^\pi=0^+_f||\hat{P}^{\dag}_{N-1}\hat {\cal P} \hat{P}_{N}+ \hat{P}^{\dag}_{N+1}\hat {\cal P} \hat{P}_{N+2}|| N, L^\pi=0^+_i \rangle \right|^2
\end{equation} 
for $(p,t)$ reactions, and 
\begin{equation}
  \label{eq:op_tp}
I(N,0_i^+\rightarrow N+1, 0_f^+)=  \left|\langle N+1, L^\pi=0^+_f||\hat{P}^{\dag}_{N+1}\hat {\cal P^\dagger} \hat{P}_{N}+\hat{P}^{\dag}_{N+3}\hat {\cal P^\dagger} \hat{P}_{N+2} || N, L^\pi=0^+_i \rangle \right|^2
\end{equation} 
for $(t,p)$ reactions. 
The structure of the two-neutron transfer operator implies that only connects the regular (intruder) part of the wave function of the parent nucleus with the regular (intruder) part of the daughter one, not existing, therefore, crossing terms connecting different sectors. The reason is because otherwise the operator will connect states with a different number of active protons, which is not allowed because we are dealing with a two-neutron transfer operator. In the case of IBM calculations with a single configuration, only the first term in Eqs. (\ref{eq:op_pt}) and (\ref{eq:op_tp}) will give a non-null contribution. Note that along this work we assume no scale factors in front of the operators and the same weight for the regular and the intruder contributions, therefore all the provided results are given in arbitrary units.   

Assuming the wave function of the involved nuclei written in the U(5) basis of the IBM as,
\begin{eqnarray}
\Psi(0_k^+;N) &=& \sum_{n_d,\tau,n_\Delta} a^{k}_{n_d,\tau,n_\Delta}[N] \psi((sd)^{N}_{n_d,\tau,n_\Delta};0^+) 
\nonumber\\
&+& \
\sum_{n_d,\tau,n_\Delta} b^{k}_{n_d,\tau,n_\Delta}[N+2]\psi((sd)^{N+2}_{n_d,\tau,n_\Delta};0^+)~,
\label{eq:wf:U5b}
\end{eqnarray}
where $n_d$ corresponds to the number of $d$ bosons, $\tau$ to the boson seniority,  $n_\Delta$ to the number of $d$ boson triplets coupled to zero,  $k$ is a rank number to label the state, and [N] and [N+2] refer to the regular and the intruder sector, respectively. The two-neutron transfer intensity for the (p,t) reaction can be expressed as,
\begin{eqnarray}
  \label{eq:op_pt2}
  I(N,0_i^+\rightarrow N-1, 0_f^+)&=& \left | \sum_{n_d,\tau,n_\Delta} \sqrt{N-n_d}\; a^{i}_{n_d,\tau,n_\Delta}[N]
  a^{f}_{n_d,\tau,n_\Delta}[N-1]  \right .  \nonumber \\
  &+&\left . \sum_{n_d,\tau,n_\Delta} \sqrt{N+2-n_d} \;b^{i}_{n_d,\tau,n_\Delta}[N+2]
  b^{f}_{n_d,\tau,n_\Delta}[N+1]   \right |^2, 
\end{eqnarray}
while for the (t,p) reaction as,
\begin{eqnarray}
  \label{eq:op_tp2}
  I(N,0_i^+\rightarrow N+1, 0_f^+)&=& \left | \sum_{n_d,\tau,n_\Delta} \sqrt{N+1-n_d}\; a^{i}_{n_d,\tau,n_\Delta}[N]
  a^{f}_{n_d,\tau,n_\Delta}[N+1]  \right .  \nonumber \\
  &+&\left . \sum_{n_d,\tau,n_\Delta} \sqrt{N+3-n_d} \;b^{i}_{n_d,\tau,n_\Delta}[N+2]
  b^{f}_{n_d,\tau,n_\Delta}[N+3]   \right |^2. 
\end{eqnarray}

From these expressions, it is self-evident that as soon as the initial and final states fully lay on a single sector, either regular or intruder, but in different ones, the value of the intensity will vanish (i.e., if the parent state is fully regular and the daughter state is purely intruder, or the opposite, the intensity will vanish). On the other hand, for vibrational nuclei the intensity will also vanish when the initial and the final state have a different phonon composition due to the selection rules $\Delta n_d=0$ and $\Delta\tau=0$. In the case of well deformed nuclei the same holds as was shown in \cite{Foss07}, i.e, the ground state band is hardly connected with the quasi-$\beta$ band or not connected at all with the double-$\gamma$ one, for instance.

\section{Two-nucleon transfer intensity calculations}   
\label{sec:two-nucleon-transfer}

\subsection{IBM with configuration mixing}
\label{sec:resultsI}
In this section, Zr, Hg, and Pt isotope chains are analyzed using the IBM-CM with the parameters obtained in former IBM-CM calculations without any additional tuning.

Along this section we will present theoretical IBM-CM results concerning (t,p) two-neutron transfer intensities, as well as the excitation energies of the unperturbed regular and intruder band-heads and the regular content of the two first $0^+$ states.
In the case of ground to excited state (p,t) reactions, they are not presented due to space reasons, but the conclusions obtained from them are essentially the same than those from (t,p) reactions.


\subsubsection{Two-neutron transfer intensities in the even-even Zr isotope chain (two configurations)}
\label{sec:zr-2C}
We use an IBM-CM Hamiltonian as summarized in Section \ref{sec:formalism}, while the details of the calculations can be found in  Ref.~\cite{Garc19}. In that work, the systematics of the spectroscopy of the low-lying collective states for Zr isotopes was assessed using IBM-CM from $A=94$ to $A=110$. For each of the isotopes, a set of parameters was fixed to reproduce excitation energies as well as E2 transition probabilities. The obtained Hamiltonians also provide a rather good description of other observables such as two-neutron separation energies, $\rho^2(E0)$ values or isotopic shifts, which point to an accurate description of the wave functions of the nuclei under study. In particular, the rapid onset of deformation when passing from $^{98}$Zr to $^{100}$Zr is well reproduced. The aim here is to take advantage of those calculations without any extra fitting and to obtain the systematic behaviour of the (t,p) two-neutron transfer intensities in this isotope chain.

\begin{figure}[htb]
\centering
\includegraphics[width=.6\linewidth]{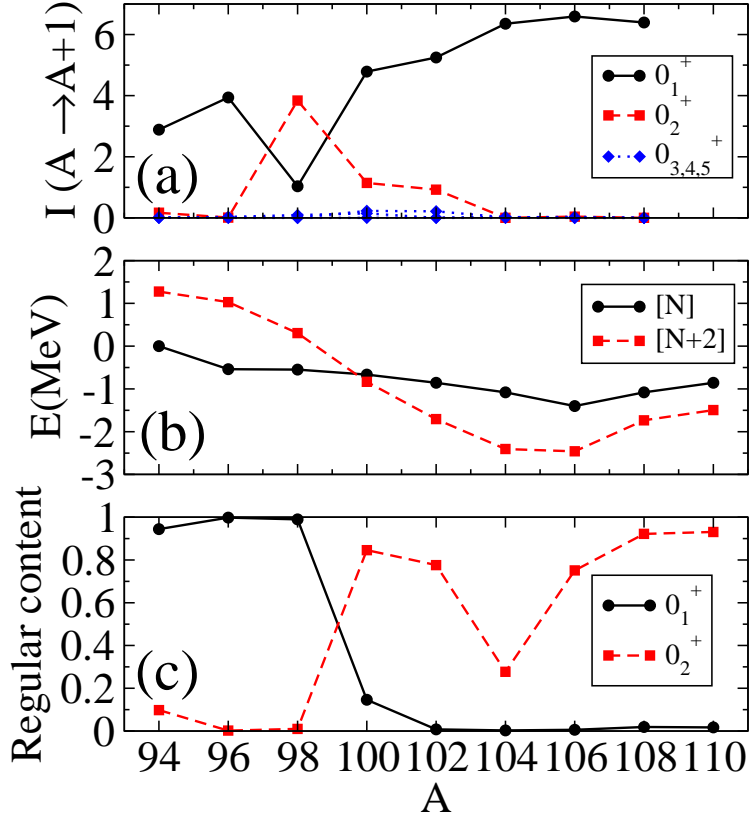}
\caption{(t,p) transfer intensities from $0_1^+$ in the parent nucleus to $0_i^+$ in the daughter for Zr isotopes given in arbitrary units using the IBM-CM Hamiltonian provided in \cite{Garc19}. In panel (a) it is depicted the value of the (t,p) transfer intensity. 
  Panel (b) corresponds to the unperturbed energy of the [N] and [N+2] band-heads. Panel (c) shows the regular content of the states $0_1^+$ and $0_2^+$ in each isotope.}
\label{fig:Zr-tp-mixing}
\end{figure}

In Fig.\ \ref{fig:Zr-tp-mixing}, we present for the Zr isotopes the obtained values for the (t,p) two-neutron transfer intensities from the ground state of the parent nucleus into the first five low-lying $0^+$ states of the daughter one (panel (a)) using the operator (\ref{eq:op_tp}) and the parameters given in Ref.~\cite{Garc19}. In order to clarify what is going on, the energies of the unperturbed $0_1^+$ regular ([N]) and $0_2^+$ intruder ([N+2]) states are presented in panel (b). In addition,  the regular content (fraction of the wave function in the regular sector, [N]) of the $0^+_1$ and $0_2^+$ states is shown in panel (c) (see \cite{Garc19} for details).

Some comments in relation to Fig.\ \ref{fig:Zr-tp-mixing} are in order.  In panel  (a) one can see that the value of the intensities to the $0_1^+$ and the $0_2^+$ states cross at the point $A=98$, precisely where the regular and the intruder configurations also cross (panel (b)). This latter fact is also manifested in the interchange of the regular content of the states $0_1^+$ and $0_2^+$ (panel (c)). The relevant observation is that at this point (between $A=98$ and $A=100$) $I(0_1^+ (A)\rightarrow 0_1^+ (A+2))<I(0_1^+ (A)\rightarrow 0_2^+ (A+2))$ while the opposite holds for the rest of cases. This fact is tightly connected  with the use of two configurations as it will be seen in section \ref{sec:zr-1C} where only a single configuration is considered.
To understand why the intensity into states other than the ground state is roughly zero, we have to resort to the argument given at the end of section \ref{sec:formalism} where we have seen that as the structure of the involved states in parent and daughter nuclei are different (one normal and the other intruder) the intensities vanish. The cancellation of the transfer intensity happens for states that either belong to different sectors (regular or intruder), for those with different phonon structure, e.g., different number of phonons in a vibrational nucleus, or having phonons of different nature in a well-deformed one. In fact, the (t,p) transfer intensity is always zero for states other than $0_{1,2}^+$, i.e., the intensity is barely fragmented.
\begin{figure}[htb]
\centering
\includegraphics[width=.9\linewidth]{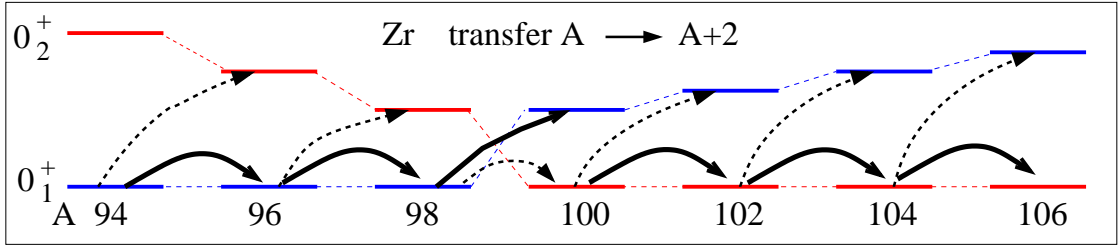}
\caption{Schematic representation for Zr isotopes of the relative position of the two first $0^+$ states and of the (t,p) transfer intensities connecting them (the width of the arrow is proportional to the value of the intensity). States with the same color and connected with dotted lines own the same structure.}
\label{fig:Zr-schematic}
\end{figure}

In Fig.~\ref{fig:Zr-schematic}, we depict schematically how the (t,p) two-neutron transfer operator connects states with similar structure and how they cross for $A=98-100$ assuming that states with a similar structure (same color), are strongly connected. What is shown in Fig.~\ref{fig:Zr-tp-mixing}a is just the manifestation of the schematic configuration crossing represented in Fig.~\ref{fig:Zr-schematic}. All along the isotope chain, two configurations coexist and they cross between $A=98$ and $A=100$. Thus, the transfer is large between vibrational ground states for $A<98$ (blue lines), and between deformed ones for $A>100$ (red lines). However, at the crossing point, i.e.,  between $A=98$ and $A=100$, the corresponding ground states have different shape (structure), spherical in $A=98$ and deformed in $A=100$, and consequently there is a drop down in the intensity of the two-nucleon transfer intensity between the ground states.

It is worth to mention a two-level mixing calculation for light Zr isotopes \cite{Fort19}, namely $^{90-96}$Zr, where the mixing amplitude is extracted considering two-neutron transfer intensities, concluding that the mixing is moderated in $^{90-94}$Zr while small in $^{96}$Zr, which is in agreement with the results presented here.

\subsubsection{Two-neutron transfer intensities in the even-even  Hg isotope chain (two configurations)}
\label{sec:hg-2C}
The nuclear region around Pb isotopes is known by the coexistence of low-lying states with different deformations \cite{heyde11}. In this region, the structure of the ground state of the isotope chains presents a rather smooth evolution, however the structure of the states as a function of the excitation energy changes abruptly, especially around midshell, i.e., N$\approx 104$, which is the result of the presence of intruder states corresponding to 2p-2h or even 4p-4h excitations across Z=82 shell closure \cite{andrey02}. The Hg isotopic chain is a  paradigmatic example of shape coexistence, with the presence of a family of intruder low-lying states and it has been studied systematically using the IBM-CM Hamiltonian as given in Ref.~\cite{Garc14b,bree14,kasia19}, obtaining a rather satisfactory description of excitation energies, $B(E2)$ values, isotopic shifts, and $\rho^2(E0)$ values. Here, again, the obtained parameters from that study allow to generate the wave functions of the different nuclei without any extra fitting.

\begin{figure}[htb]
\centering
\includegraphics[width=.6\linewidth]{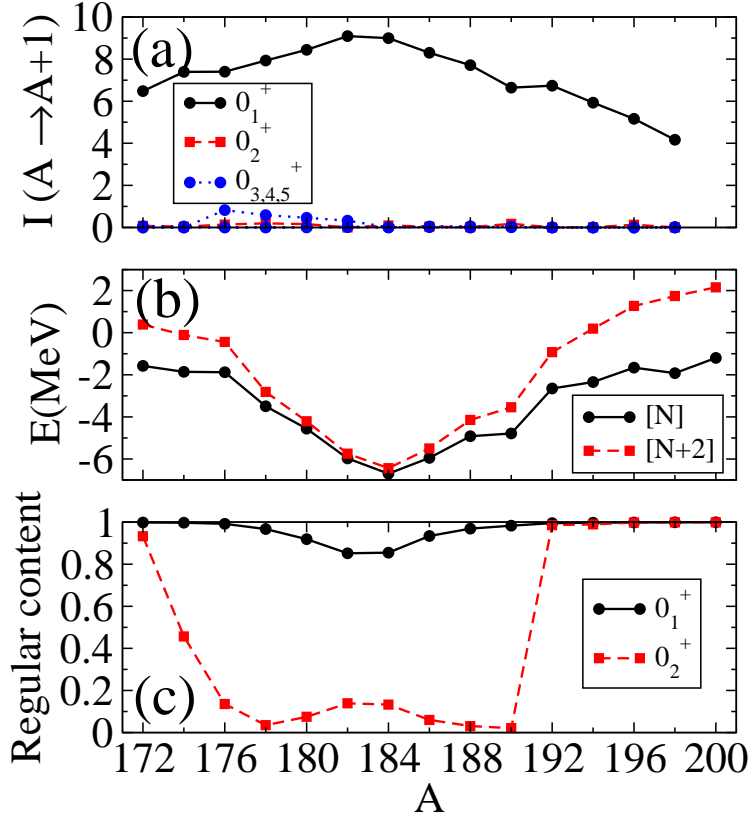}
\caption{The same as Fig.~\ref{fig:Zr-tp-mixing} but for Hg isotopes. IBM-CM results taken from \cite{Garc14b}.}
\label{fig:Hg-tp-mixing}
\end{figure}

In Fig.~\ref{fig:Hg-tp-mixing}, the value of the calculated (t,p) transfer intensities for the Hg isotope chain between the ground state of the parent and the first five states of the daughter nucleus is presented in panel (a), in  panel (b) the behaviour of the energies obtained for the unperturbed [N] and [N+2] lowest $0^+$ states is plotted, finally, the regular content of the states $0^+_1$ and $0_2^+$ are depicted in panel (c) (see \cite{Garc14b} for details).

\begin{figure}[htb]
\centering
\includegraphics[width=.6\linewidth]{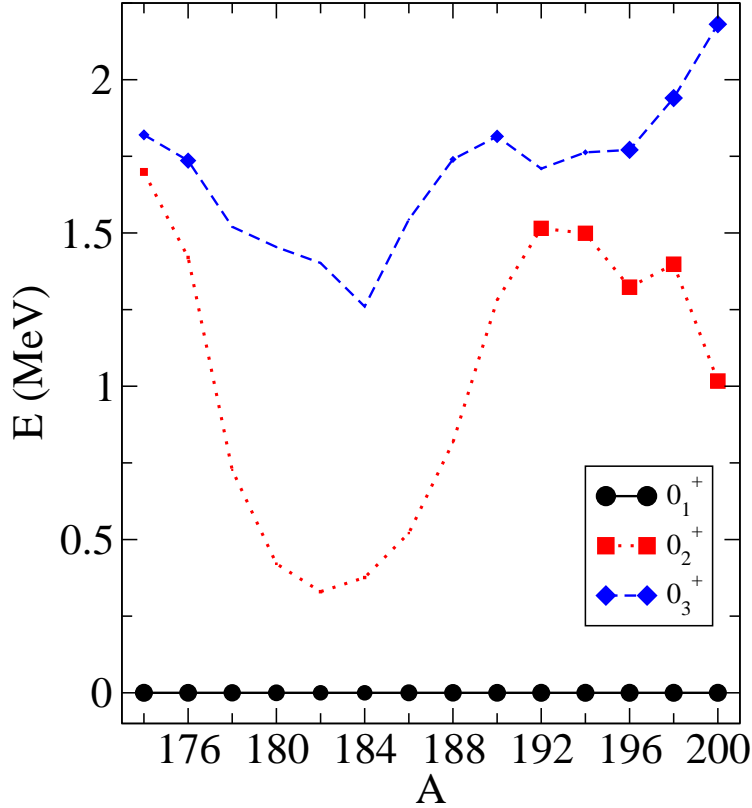}
\caption{Energy systematics for the low-lying $0^{+}$ states in the Hg isotopes. The size of the dot is proportional to the regular component of the wave function.}
\label{fig:Hg-energy-systematics}
\end{figure}

For this isotope chain, the two competing configurations never cross, as can be observed in Fig.~\ref{fig:Hg-tp-mixing}b, with the almost pure [N] configuration being always below the pure [N+2] one. Because of that, it is shown that there is little mixing between both configurations all along the isotope chain and, consequently, the dominant two-neutron transfer intensity for all the isotopes is between the ground states, i.e., $I(0_1^+ (A)\rightarrow 0_1^+ (A+2))$. The transfer to any other low-lying $0^+$ is very small for all isotopes. We conclude that in Hg isotopes it is not observed any sizeable effect in the transfer intensities involving $0^+$ states because there is almost no mixing between the regular and the intruder sector (panel (c)). It is seen that the  $0_2^+$ state in isotopes $A=172$ and $A=192-200$ is mostly the second regular $0^+$ state, i.e. it belongs to the [N] configuration, while from $A=174$ to $190$ the lowest intruder $0^+$ state belonging to the $[N+2]$ configuration comes lower than the second $0^+$ of $[N]$, however, the intensity still vanish because ground and $0_2^+$ states present a different number of vibrational phonons. 
To see more clearly this fact, we plot in \ref{fig:Hg-energy-systematics} the energy systematics of the first three $0^+$ states together with the value of the regular component.

\subsubsection{Two-neutron transfer intensities in the even-even Pt isotope chain (two configurations)}
\label{sec:pt-2C}
In the Pb region another very interesting isotope is Pt. The Pt isotopes also present low-lying intruder states coming from the $[N+2]$ configuration. The systematics of this isotope chain including configuration mixing has been studied within the IBM-CM Hamiltonian in Refs.~\cite{Garc09,Garc11}. One of the main conclusion from these works is that in Pt a very large degree of mixing between the intruder and the regular states exists and that shape coexistence is somehow concealed, with barely differences between the calculations with one or two configurations. The parameters obtained from these works were used to generate the wave functions of the relevant states considered in the present work. 

\begin{figure}[htb]
\centering
\includegraphics[width=.6\linewidth]{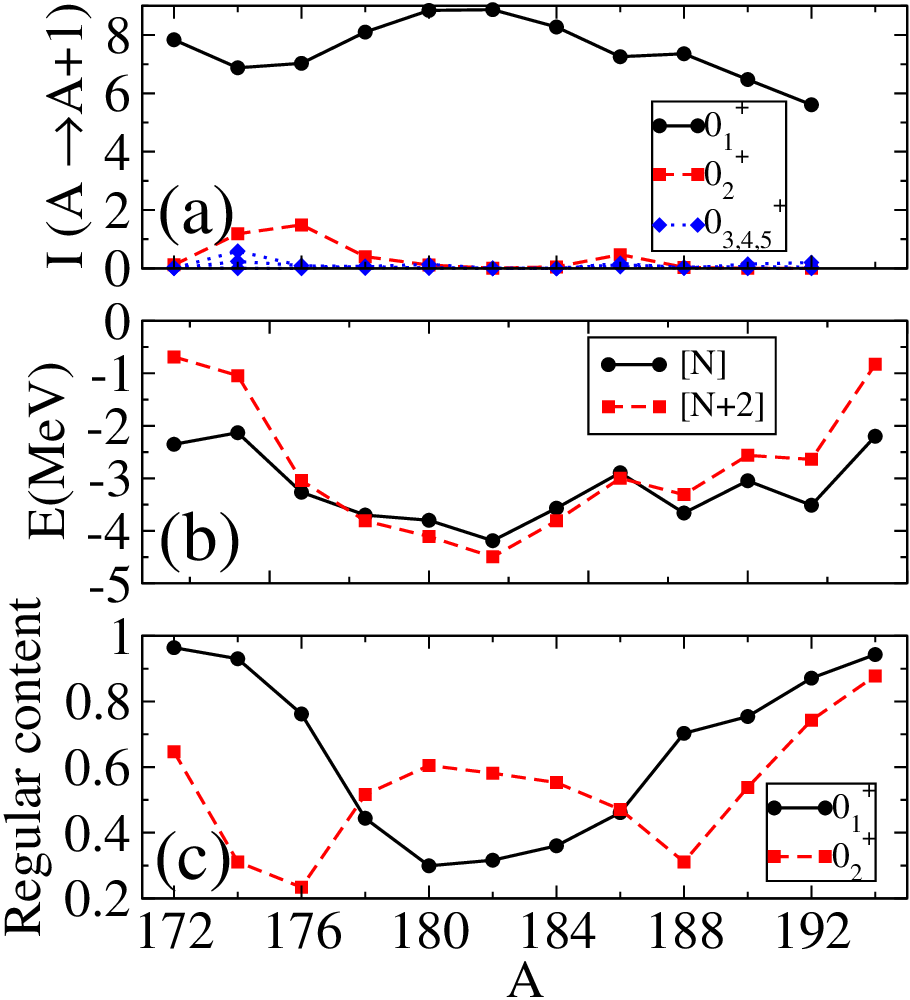}
\caption{The same as Fig.~\ref{fig:Zr-tp-mixing} but for Pt isotopes. IBM-CM results taken from \cite{Garc09}.}
\label{fig:Pt-tp-mixing}
\end{figure}
In Figure \ref{fig:Pt-tp-mixing}, the calculated values for (t,p) transfer intensities from the $0^+_1$ in the parent nucleus to $0^+_i$ states in the daughter one for the Pt isotope chain are plotted in panel (a). As in the case of Hg, in panel (b), the energies of the unperturbed $[N]$ and $[N+2]$ $0^+$ band-heads are presented and in panel (c), the regular content of the states $0^+_1$ and $0_2^+$ are shown. Panel (a) is very similar to the Hg case, but looking at panel (b) and (c), one notices  important differences. In the Pt isotopes both relevant configurations compete, cross and stay very close for most of the isotopes. Looking at panel (c), it can be seen that both the $0_1^+$ and the $0_2^+$ states are strongly mixed, around $50\%$ in both of them for many midshell isotopes. Then, the systematics of the two-neutron transfer intensity (panel (a)) shows that the transfer between the ground states in neighbouring isotopes dominates. However, in this case it does not follow that the $[N]$ configuration dominates because it is strongly mixed all the way with the intruder $[N+2]$ configuration. In this case, the intruder and the regular configurations cross before and after the midshell, becoming the intruder configuration the ground state around midshell.
Concerning the two-neutron transfer intensities, one notices some effect at the place where the states cross, but it is hardly noticeable and $I(0_1^+ (A)\rightarrow 0_1^+ (A+2))$ remains the dominant intensity all the way. This fact could be considered as unexpected because of the crossing of the configuration and of the large mixing between them, roughly $50\%$, at the points where the configurations cross. Indeed, the $50\%$ mixing in both the father and daughter isotopes allows to get a large fraction of intensity from both sectors. Note that the transition operator connect the regular (intruder) sector of the father nucleus with the regular (intruder) one in the daughter nucleus, therefore both sectors contribute but either in a constructive (for $0^+_1 \rightarrow 0^+_1$) or destructive (for $0^+_1 \rightarrow 0^+_2$) way. The leading transition remains quite stable and only a minor lowering (an a modest increase in the transfer to the $0_2^+ (A+2)$ state) is observed around the crossing points. According to the results, the contributions for the transition to $0_1^+$ sum up in a constructive form while in a negative way for the transition to the $0_2^+$ state that is almost zero all the way.

\subsection{IBM with a single configuration Hamiltonian}
\label{sec:resultsII}
In this section the two-neutron transfer intensity for Zr, Pt, and Sm isotopes are explored using the IBM with a single configuration. As in previous section, the parameters of the Hamiltonians are fixed in previous studies and here the wave functions are used without further tuning.

Along this section we will present theoretical IBM results using a single configuration concerning (t,p) two-neutron transfer intensities, as well as the excitation energy of the first $0^+$ excited state and the E2 reduced transition probability between the $0_2^+$ and $2_1^+$ states. The two latter observables are considered as hints for the existence of a QPT \cite{Cejn09,Cejn10}.

\subsubsection{Two-neutron transfer intensities in the even-even Zr isotope chain (single configuration)}
\label{sec:zr-1C}
In order to study the two-neutron transfer intensities based on systematic calculations within the Zr isotope chain using a single configuration, we use here the IBM Hamitonian and the parameters obtained in Ref.~\cite{Garc05}, without  extra fitting for the calculation of the two-neutron transfer intensity. In Ref.~\cite{Garc05} the spectroscopic properties of even-even Zr isotopes were studied in detail with the goal of an appropriate reproduction of the two-neutron separation energy. Moreover, this work pointed to the existence of a QPT being $^{100}$Zr the critical nucleus.  

\begin{figure}[htb]
\centering
\includegraphics[width=.6\linewidth]{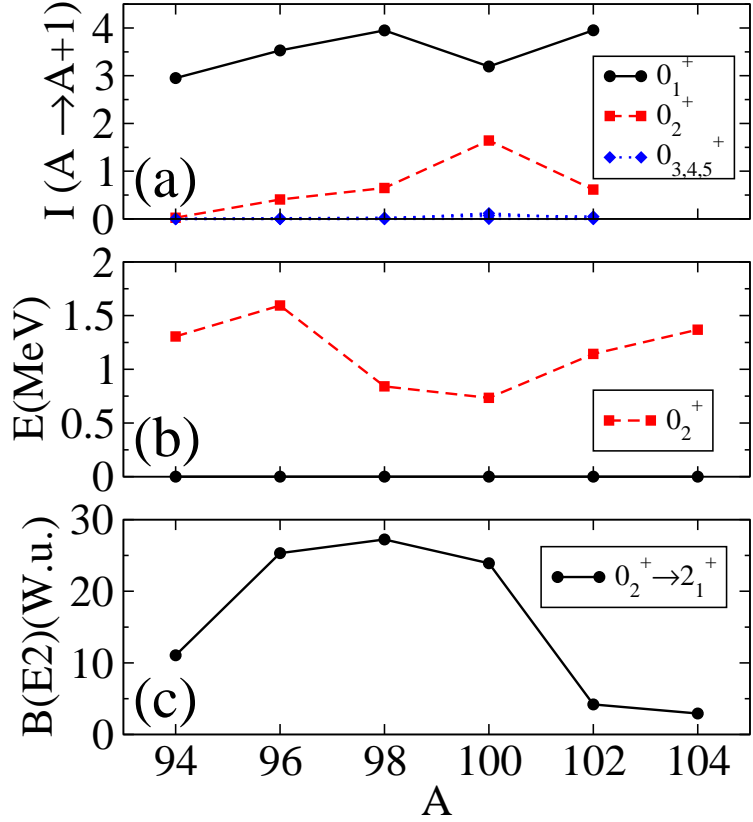}
\caption{(a) (t,p) transfer intensities for Zr isotopes from $0^+_1$ in the parent nucleus to $0^+_i$ in the daughter one, given in arbitrary units using the IBM with a single configuration and the parameters from \cite{Garc05}. (b) Excitation energy of the state $0_2^+$. (c) $B(E2:0_2^+\rightarrow 2_1^+)$ values.}
\label{fig:Zr-tp-single}
\end{figure}

In  panel (a) of Fig.~\ref{fig:Zr-tp-single}, (t,p) two-neutron transfer intensities from $0^+_1$ in the parent nucleus to $0^+_i$ in the daughter one for Zr isotopes, described using a single configuration calculation, are shown. As complementary observables, in panel (b) the excitation energy of the $0_2^+$ state is plotted, and in panel (c) the $B(E2:0_2^+\rightarrow 2_1^+)$ values are presented. Regarding the systematics of the intensities, one notices a certain dropping and an associated rising in  $I(0_1^+ (A)\rightarrow 0_1^+ (A+2))$ and $I(0_1^+ (A)\rightarrow 0_2^+ (A+2))$, respectively, at $A=100$, while the transfer to other $0^+$ states remains all the way almost at zero value, which supposes little fragmentation of the strength even at $A=100$ where a QPT is supposed to exist.  
Concerning the $E(0_2^+)$ excitation energy (panel (b)), a minimum for $A=100$ is observed, however the energy of the experimental one at $A=100$ is much lower because it corresponds to a  state of vibrational nature that is not considered in our IBM calculation that use a single configuration. In the present calculation, owing to the use of a single configuration, both ground and excited $0_2^+$ states (intruder) have a deformed character. Therefore, the presented energy systematics is much smoother than the experimental one. The lowering of this excitation energy with a minimum for $^{100}$Zr is consistent with the presence of a critical point at $A=100$.

In the case of $B(E2:0_2^+\rightarrow 2_1^+)$ (panel (c)) the situation is analogous and the dropping down when passing from $A=100$ to $A=102$ is much smoother than the experimental observation. This dropping arises from the passing of a spherical shape, where the transition involves passing of a two-phonon into a one-phonon state, to a deformed one, where the transition implies the connection of states belonging to different irreps (if they were in the SU(3) limit). When calculating this same observable using two configurations, instead of modifying the Hamiltonian as a function of the neutron number to generate the observed sudden change in deformation (see \cite{Garc05}), the onset of deformation is generated through the crossing of two families, one spherical and the other deformed (see Section \ref{sec:zr-2C}) and, in general, all the analyzed quantities present a much faster rate of change.  

Note that the theoretical values for $B(E2:0_2^+\rightarrow 2_1^+)$ have been calculated with an effective charge, $e=2.8 \sqrt{\text{W.u.}}$ which has been fixed to reproduce the experimental value of $B(E2:2_1^+\rightarrow 0_1^+)$ \cite{Garc05}. This observable resembles the behavior of the order parameter of a QPT, with a null value in one of the phases and a rapid increase when passing to the other. The observed behaviour is much smoother than the experimental one and than that obtained using two configurations (see \cite{Garc19}). Clearly the smoother trends, obtained theoretically in this subsection are due to the use of a single configuration. This IBM calculation (with a single configuration) was tailored to reproduce the rapid changes observed around $^{100}$Zr, such as two-neutron separation energies, $E(2_1^+)$, and $E(4_1^+)/E(2_1^+)$, however the trend for $E(0_2^+)$ cannot be correctly reproduced, in particular for $^{100}$Zr, because in this case the $0_2^+$ state corresponds to a regular state while the ground state to a intruder one. In other words, the passing from a spherical to a deformed shape has been generated changing appropriately the parameters of the Hamiltonian but when regular and intruder states are involved simultaneously in the description of a given observable, it is not possible to provide an accurate description using a single configuration.

All in all makes that the (t,p) transfer intensity changes are much smoother for single configuration calculations than in the two mixing-configuration case. Moreover, for all the isotopic chain  $I(0_1^+ (A)\rightarrow 0_1^+ (A+2))>I(0_1^+ (A)\rightarrow 0_2^+ (A+2))$ in the single configuration calculation, while for the configuration mixing study 
$I(0_1^+ (A=98)\rightarrow 0_1^+ (100))<I(0_1^+ (98)\rightarrow 0_2^+ (100))$ reflecting the crossing of two configurations (see Fig.~\ref{fig:Zr-tp-mixing}).

\subsubsection{Two-neutron transfer intensities in the even-even Pt isotope chain (single configuration)}
\label{sec:pt-1C}
In this section, we study the systematics of the two-neutron transfer intensity of even-even Pt isotopes using the IBM Hamiltonian with a single configuration as obtained in \cite{cutcham05}. In this reference the excitation energy systematics and the E2 transition rates of the even-even $^{172-196}$Pt isotopes where nicely described. Here, we generate the wave functions and study the two-neutron transfer without any additional parameter fitting.  

\begin{figure}[htb]
\centering
\includegraphics[width=.6\linewidth]{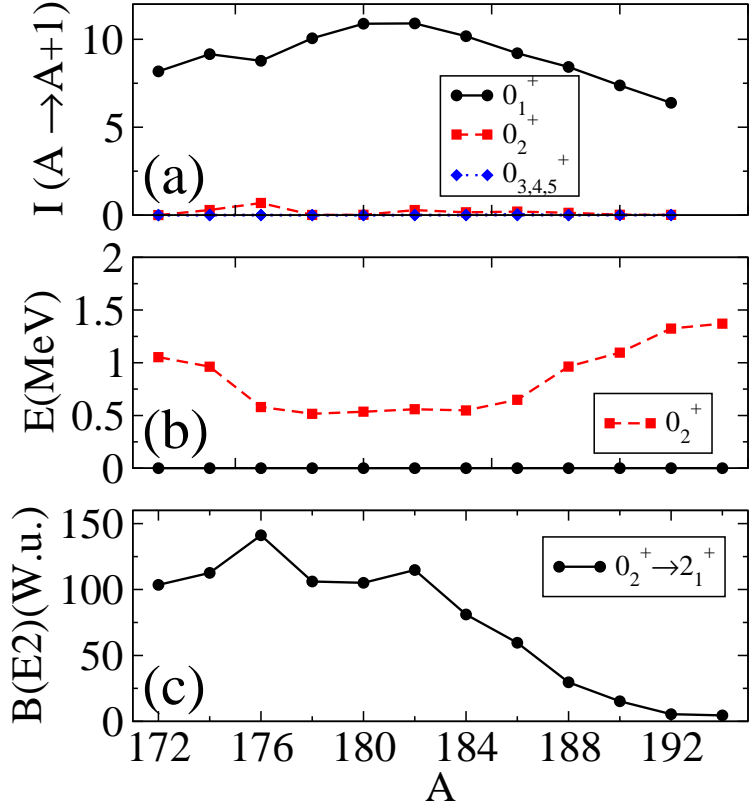}
\caption{The same as Fig.~\ref{fig:Zr-tp-single} but for Pt isotopes. IBM results taken from \cite{cutcham05}.}
\label{fig:Pt-tp-single}
\end{figure}

In panel (a) of Fig.~\ref{fig:Pt-tp-single}, (t,p) two-neutron transfer intensities in Zr isotopes obtained with a single configuration calculation are shown. As complementary observables, in panel (b) the excitation energy of $0_2^+$ is plotted, and in panel (c) the $B(E2:0_2^+\rightarrow 2_1^+)$ values are depicted. Regarding the systematics of the intensities, one notices a relatively constant and large value of  $I(0_1^+ (A)\rightarrow 0_1^+ (A+2))$, being the rest of intensities almost zero except $I(0_1^+ (A)\rightarrow 0_2^+ (A+2))$ at $A=176$. 
The excitation energy of the $0_2^+$ state depicted in panel (b) shows a rather constant value not being affected by the presence of the midshell at $A=182$. The structure of the state is evolving along the isotope chain, though slowly, as can be readily seen in the continuous dropping of the $B(E2:0_2^+\rightarrow 2_1^+)$ value (panel (c)), however there is almost no influence in the value of the (t,p) transfer intensity.

The Hamiltonian obtained in \cite{cutcham05} provides a very good and detailed description of the spectroscopic properties of the whole isotope chain, almost as good as using IBM-CM \cite{Garc09}. Besides, the values of the two-neutron intensities provided by both approaches is almost equivalent, without any major difference.

\subsubsection{Two-neutron transfer intensities in the even-even Sm isotope chain (single configuration)}
\label{sec:sm-1C}
The Sm isotope chain is considered as a crystal-clear example of a QPT from spherical to axially deformed shapes at $N=90$. There are many indications of abrupt changes in this isotope chain: the two-neutron separation energies, B(E2)'s values, energy ratios, etc. In Ref.~\cite{Foss02} systematic calculations for a large set of isotopes, among them for Sm, have been carried out using IBM with a single configuration. In this section we will use the IBM parameters for the case of even-even Sm isotopes without any tuning to generate the corresponding wave functions. With those, the two-neutron transfer intensities between $0^+$ states in the initial and final nuclei have been calculated. To compute B(E2) values we have considered the same effective charge for the whole chain, $e_{eff}=2.2 \sqrt{\text{W.u.}}$ (this value of the effective charge reproduces the experimental value $B(E2:4_1^+\rightarrow 2_1^+)$ in $A=152$).

\begin{figure}[htb]
\centering
\includegraphics[width=.6\linewidth]{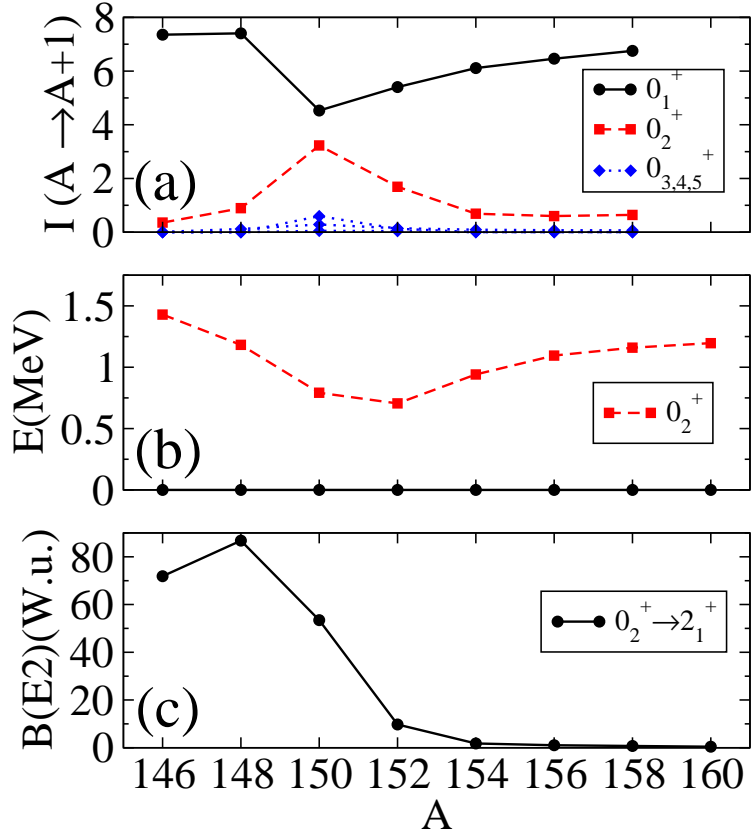}
\caption{The same as Fig.~\ref{fig:Zr-tp-single} but for Sm isotopes. IBM results taken from \cite{Foss02}.}
\label{fig:Sm-tp-single}
\end{figure}

In Figure \ref{fig:Sm-tp-single}, the value of the calculated (t,p) transfer intensities from the $0^+$ ground state into the first five low-lying $0^+$ states in the daughter nucleus have been plotted in panel (a). In panel (b), the systematics of the excitation energies of the $0_2^+$ states are presented. Finally, the $B(E2:0_2^+\rightarrow 2_1^+)$ values are depicted in panel (c). Note in panel (b) that the drop down of the $0_2^+$ energy is somehow similar to what happens in the case of the presence of intruder states, though the minimum is not placed at midshell, but where a QPT is supposed to exist.
Note that rare earth region is known by the interplay between quadrupole degrees of freedom and pairing vibrations as recently studied in \cite{Xian20}. Taking into account that in \cite{Foss02} a phenomenological IBM Hamiltonian has been used to describe the excitation energies and B(E2) transition rates, it is expected that quadrupole and pairing degrees of freedom have been correctly incorporated.
One observes in panel (c) the abrupt change of $B(E2:0_2^+\rightarrow 2_1^+)$ from a reasonable large value to basically zero  when passing from $A=150$ to $A=154$ which resembles the behaviour of an order parameter in a QPT. 
Concerning the (t,p) intensity values presented in the panel (a), a sudden increase of $I(0_1^+ (A)\rightarrow 0_2^+ (A+2))$ accompanied by a drop down of $I(0_1^+ (A)\rightarrow 0_1^+ (A+2))$  is observed at $A=150 \rightarrow 152$ ($N=90$). The two-neutron transfer intensities to $0_3^+$, $0_4^+$, and $0_5^+$ in the daughter nucleus are very small. Note that all the way $I(0_1^+ (A)\rightarrow 0_1^+ (A+2))>I(0_1^+ (A)\rightarrow 0_2^+ (A+2))$. The early IBM calculation carried out in \cite{Saha79} provides almost identical results. 

What we see in this case is similar to what was observed in the IBM calculation with a single configuration for Zr, though in the case of Sm the intensities to the states $0^+_{3,4,5}$ were a little stronger than for Zr. Therefore, in Sm there is certain degree of fragmentation of the (t,p) strength. 

Very recently a two-level mixing calculation has been carried out for $^{150-152}$Sm isotopes \cite{Fort19b} trying to reproduce two-neutron transfer intensities and E2 transition rates, concluding that it exits a moderate mixing between two families of states.

\section{Conclusions}
\label{sec:conclusions}
In this work, we have calculated the value of the (t,p) two-neutron transfer intensities taking advantage of previous systematic studies using the IBM with and without configuration mixing in the regions of Zr, Pb, and Sm which provide the corresponding wave functions of the $0^+$ low-lying states in each isotope chains and then allow to calculate, with no additional tuning, the value of the intensity. Our main goal was to check whether or not the two-neutron intensity is a reliable observable to distinguish between the existence of shape coexistence and a QPT as previously suggested in \cite{Vitt18,Vittu18a}, where it was observed than in schematic calculations the two-neutron intensities suffer abrupt changes in both cases, but the intensity is much more fragmented in the presence of a quantum phase transition than for shape coexistence.  

The first studied chain of isotopes has been Zr, using both approaches IBM-CM and IBM with a single configuration. 
In both cases it is observed a rapid change in the (t,p) transfer intensity at $A=100$, being much more abrupt in the case of shape coexistence, but no fragmentation has been observed in the QPT case.
Clearly the existence or not of fragmentation in the case of a QPT depends on the precise characteristics of the used Hamiltonians, being fragmented in the schematic calculation shown in \cite{Vitt18,Vittu18a,Foss07} but not in the  realistic case shown here.  

The second case has been the Hg isotope chain which has been studied only using IBM-CM \cite{Garc14b} because using IBM with a single configuration it is not possible to describe the experimental spectroscopic systematics. In this case there is no sizeable effect of shape coexistence on the value of the intensities because the intruder configuration is always well above the regular one and they are hardly mixed.

The third case corresponds to the Pt isotope chain that has been analyzed using both the IBM-CM \cite{Garc09,Garc11} and the IBM with a single configuration \cite{cutcham05}. 
The obtained systematics is somehow surprising due to the large mixing before and after the midshell, that would be expected to strongly affect the two-neutron intensities into the $0_1^+$ and $0_2^+$ states of the daughter nuclei, however the particular phases of the different components of the wave function produce a constructive effect for the $0_1^+$ state, while destructive for $0_2^+$.

The final case is the chain of Sm isotopes which has been studied using an IBM Hamiltonian with a single configuration \cite{Foss02}.
Clearly, the occurrence of a QPT has a strong influence on the two-neutron intensity, affecting to the transitions into the $0_1^+$ and $0_2^+$ states but without any further fragmentation. This is in disagreement with the conclusion raised in \cite{Foss07,Vitt18,Vittu18a} where a schematic QPT was simulated and a large fragmentation was observed, which suggests that the results should be sensitive to the precise characteristic of the Hamiltonian evolution along the isotope chain.

The evolution of nuclear structure is the result of the fine balance of the nuclear interaction that can be schematically understood in terms of the competition between the quadrupole interaction, that tends to deform the nucleus, and the monopole part, which is very much dependent on the specific orbitals around the shell closures, that tends to keep the nucleus as spherical. This competition is at the origin of the appearance of deformation in the nuclear chart. In the case of Zr and Sm isotope, where it appears a rapid onset of deformation, the competition can be easily understood in terms of the effect suggested by Federman and Pittel in 1979 \cite{Fede79b} where they emphasized the importance of the simultaneous filling of neutrons and protons spin-orbit partners, $\pi g_{9/2}$ and $\nu g_{7/2}$ in the case of Zr and $\pi h_{11/2}$ and $\nu h_{9/2}$ in the case of Sm, to understand the appearance of deformation. In the case of Zr, the description of the spectroscopic properties need the inclusion of a intruder configuration while not in the case of Sm, though in both cases there is a rapid onset of deformation and several states present a quite different degree of deformation (see \cite{Togashi16} for Zr and \cite{Iach98} for Sm). In both cases there is a strong influence on the two-neutron transitions, but the precise trend along the chain depends on the precise details of the interaction and it is not possible to clearly disentangle the shape coexistence picture from the QPT one only using the two-neutron transfer intensity.

\section{Acknowledgment}
One of the author (JEGR) thanks K.~Heyde for enlightening discussions. This work has been partially supported by the Ministerio de Ciencia e Innovaci\'on (Spain) under projects number FIS2017-88410-P, PID2019-104002GB-C21 and PID2019-104002GB-C22, by the Consejer\'{\i}a de Econom\'{\i}a, Conocimiento, Empresas y Universidad de la Junta de Andaluc\'{\i}a (Spain) under Group FQM-160 (JMA) and FQM-370 (JEGR), by the European Regional Development Fund (ERDF), ref.\ SOMM17/6105/UGR, and by the European Commission, ref.\ H2020-INFRAIA-2014-2015 (ENSAR2). Resources supporting this work were provided by the CEAFMC and the Universidad de Huelva High Performance Computer (HPC@UHU) funded by ERDF/MINECO project UNHU-15CE-2848.
\bibliography{references-IBM-CM,references-QPT}
\end{document}